\documentclass[preprint, 3p, twocolumn, longtitle]{elsarticle}
\usepackage[utf8]{inputenc}
\usepackage{graphicx}
\usepackage{booktabs}
\graphicspath{{good_figures/}}
\usepackage{caption}
\usepackage{tikz}
\usepackage{url}
\usepackage{comment}
\usepackage{epstopdf}
\newcommand{\U}[1]{\ensuremath{\mathrm{\ #1}}}
\newcommand{\UU}[2]{\ensuremath{\mathrm{\ #1^{#2}}}}

\journal{Astroparticle Physics}

\begin{document}

\begin{frontmatter}

\title{Detection of the Crab Nebula with the 9.7 m Prototype Schwarzschild-Couder Telescope}

\author[add0]{C.~B.~Adams}
\author[add1]{R.~Alfaro}
\author[add2]{G.~Ambrosi}
\author[add3]{M.~Ambrosio}
\author[add3]{C.~Aramo}
\author[add4]{T. ~Arlen  }
\author[add5]{P.~I.~Batista}
\author[add6]{W.~Benbow}
\author[add2,add7]{B.~Bertucci}
\author[add8,add9]{E.~Bissaldi}
% ...

\address[add0]{Physics Department, Columbia University, New York, NY 10027, USA}
\address[add1]{Instituto de F\'{i}sica, Universidad Nacional Aut\'{o}noma de M\'{e}xico, Ciudad de M\'{e}xico, Mexico}
\address[add2]{INFN Sezione di Perugia, 06123 Perugia, Italy}
\address[add3]{INFN Sezione di Napoli, 80126 Napoli, Italy}
\address[add4]{Department of Physics and Astronomy, University of California, Los Angeles, CA 90095, USA  }
\address[add5]{Deutsches Elektronen-Synchrotron, Platanenallee 6, 15738 Zeuthen, Germany}
\address[add6]{Center for Astrophysics | Harvard \& Smithsonian, Cambridge, MA 02138, USA}
\address[add7]{Dipartimento di Fisica e Geologia dell’Universit\`a degli Studi di Perugia, 06123 Perugia, Italy}
\address[add8]{Dipartimento Interateneo di Fisica dell’Universit\`a e del Politecnico di Bari, 70126 Bari, Italy}
\address[add9]{INFN Sezione di Bari, 70125 Bari, Italy}
% ...

\cortext[cor1]{Corresponding authors}


\begin{abstract} \label{Abstract}
    The Schwarzschild-Couder Telescope (SCT) is a telescope concept proposed for the Cherenkov Telescope Array. It employs a dual-mirror optical design to remove comatic aberrations over an $8^{\circ}$ field of view, and a high-density silicon photomultiplier camera (with a pixel resolution of 4 arcmin) to record Cherenkov emission from cosmic ray and gamma-ray initiated particle cascades in the atmosphere. The prototype SCT (pSCT), comprising a 9.7~m diameter primary mirror and a partially instrumented camera with 1536 pixels, has been constructed at the Fred Lawrence Whipple Observatory. The telescope was inaugurated in January 2019, with commissioning continuing throughout 2019. We describe the first campaign of observations with the pSCT, conducted in January and February of 2020, and demonstrate the detection of gamma-ray emission from the Crab Nebula with a statistical significance of $8.6\sigma$.
\end{abstract}


\end{frontmatter}

\section{Introduction}
The imaging atmospheric Cherenkov technique was pioneered at the Whipple Observatory \cite{Weekes} and has been employed over the past two decades by VERITAS \cite{VERITAS}, MAGIC \cite{MAGIC} and H.E.S.S. \cite{HESS} with remarkable success. Over two hundred astrophysical sources of $>30\U{GeV}$ gamma-ray emission have been measured\footnote{http://tevcat.uchicago.edu/}, revealing relativistic particle acceleration processes in a variety of Galactic and extragalactic environments. 
The Cherenkov Telescope Array (CTA) \cite{CTA} is an observatory currently under development and construction which will explore this gamma-ray sky with greatly improved sensitivity, energy and angular resolution. It will consist of two large telescope arrays: one in the northern hemisphere, on the Canary Island of La Palma in Spain, and the other in the south, at Paranal in Chile. Multiple telescope designs are being implemented in order to provide sensitive coverage over at least four orders of magnitude in energy, from 30\U{GeV} to 300\U{TeV}. 

Atmospheric Cherenkov telescopes are designed to image and record the few nanosecond-duration Cherenkov emission from cosmic ray and gamma-ray initiated atmospheric particle cascades (e.g. \cite{Hillas}), leading to some unique design constraints. In particular, they require a field of view wider than a few degrees to capture all of the Cherenkov emission from air showers, which are offset from the optical axis of the telescope and can extend over a few degrees at the highest energies. A wide field of view also allows the study of astrophysical sources with large angular extent, which are relatively common in the gamma-ray regime, and improves the prospects for detecting serendipitous, or poorly located, astrophysical transient events such as gamma-ray bursts, and astrophysical neutrino and gravitational wave counterparts. This wide field of view must be coupled to a large aperture mirror in order to collect a sufficient number of Cherenkov photons to form a shower image. Both requirements (wide field of view and large aperture) push towards a small focal ratio, ideally less than f/1.0. This presents its own problems, however, since off-axis optical aberrations, particularly coma, become severe. Existing wide-field designs mitigate this to some extent by using a prime-focus Davies-Cotton approach \cite{DC}, in which a single tessellated reflector is constructed using identical mirror facets placed on a spherical surface with a radius equal to the focal length of the facets. 

The dual-mirror Schwarzschild-Couder optical design \cite{Schwarzschild, Couder}, derived from the exact Schwarzschild aplanatic solution described in detail in \cite{pSCT_optical_concept}, resolves the problem of off-axis optical aberrations more effectively by correcting comatic aberrations over a wide field. In addition, it provides a greatly reduced focal plate-scale which allows to take advantage of high density, high photon detection efficiency, silicon photosensor arrays at the focal plane, which is curved in order to reduce astigmatism. A Schwarzschild-Couder optical system has been selected for the small-sized telescopes  (SSTs) of CTA, which will be the most numerous CTA telescopes, providing sensitive coverage of the highest energy gamma-ray band up to $300\U{TeV}$. The ASTRI Project has successfully demonstrated the practicality of this design with the ASTRI-Horn SST prototype instrument \cite{ASTRI}, which has been used to detect gamma-ray emission from the Crab Nebula with a camera based on silicon-photomultiplier (SiPM) sensors \cite{ASTRICrab}. The ASTRI design has a $4.3\U{m}$ diameter primary mirror and a $1.8\U{m}$ secondary, giving an effective photon collection area (after accounting for shadowing) of approximately $5\UU{m}{2}$. 

\begin{figure}[h]
    \centering
    \includegraphics[width=\columnwidth]{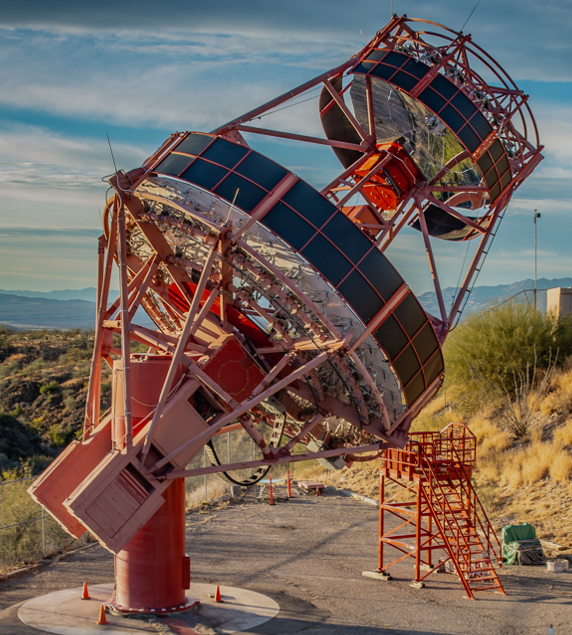}
    \captionof{figure}{The $9.7\U{m}$ prototype Schwarzschild-Couder telescope installed at the Fred Lawrence Whipple Observatory in Amado, Arizona USA. }
    \label{fig:telescope}
\end{figure}

A dual-mirror Schwarzschild-Couder design, with f/0.58, is also being investigated as a solution for the medium-sized telescopes (MSTs) of CTA, in addition to a more conventional Davies-Cotton design \cite{DC-MST}. The MSTs will provide greatest sensitivity in the core energy range of the observatory, centered around $1\U{TeV}$. The prototype of this Schwarzschild-Couder telescope  design (hereafter pSCT) is illustrated in Figure~\ref{fig:telescope}, and fully described in the following section.  The telescope was inaugurated, and recorded first light, in January 2019. Commissioning and first observations continued until February 2020, shortly before the observatory was temporarily closed due to the COVID-19 pandemic. A reasonable exposure was collected on the Crab Nebula, a bright, stable, standard gamma-ray source in the TeV regime \cite{Vacanti}. We discuss here some details of the commissioning and observations, and present a detection of gamma-ray emission from the Crab Nebula with the pSCT.

\section{The prototype Schwarzschild-Couder telescope}
\begin{figure}[h]
    \centering
    \includegraphics[width=\columnwidth]{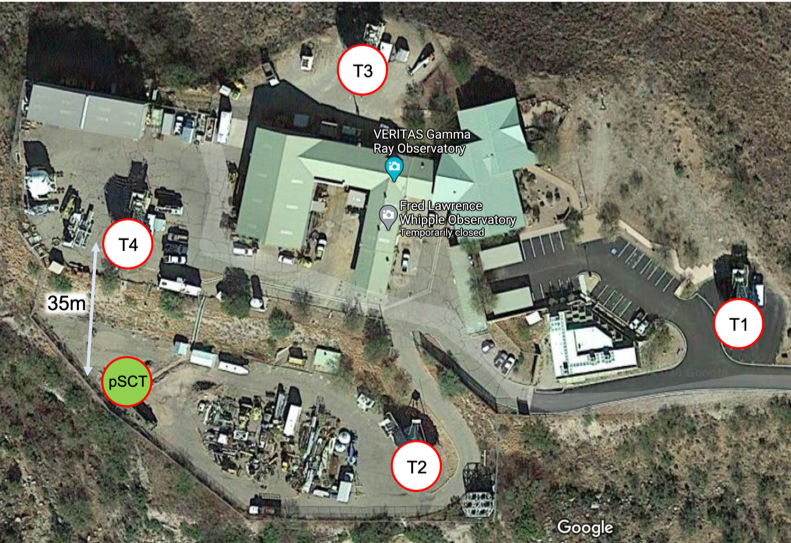}
    \captionof{figure}{A plan view of the Fred Lawrence Whipple Observatory, illustrating the locations of the four VERITAS telescopes and the pSCT. }
    \label{fig:site_layout}
\end{figure}

The pSCT is located at the Fred Lawrence Whipple Observatory basecamp in southern Arizona,  at latitude
31$^\circ$ $40^{\prime}$ $29^{\prime\prime}$ N, longitude 110$^\circ$
$57^{\prime}$ $10^{\prime\prime}$ W and at an elevation of $1270\:\mathrm{m}$ above sea level. It shares this site  with the 4-telescope VERITAS array, and occupies the position of the first VERITAS telescope, prior to its relocation in 2009. The pSCT position with respect to the basecamp buildings and the VERITAS array is shown in Figure~\ref{fig:site_layout}. 

The pSCT central support tower and telescope positioning system are adapted, with minimal changes, from the Davies-Cotton MST design \cite{DC-MST}. The mechanical design of the telescope is described in \cite{pSCT-OSS}. The telescope optical system \cite{psct_optical_spie} consists of a $9.7\U{m}$ diameter primary reflector and a $5.4\U{m}$ secondary. The primary is constructed of 48 individual mirror panels: an inner ring of 16 panels and an outer ring of 32 panels. The secondary reflector is also segmented, with 8 panels forming an inner ring and 16 forming the outer ring. The reflecting panels are based on a low-weight (approximately $10\U{kg}\UU{m}{-2}$) sandwiched substrate produced by the Italian Media Lario company via replication technology  using thin (thermally pre-shaped) glass foils separated by an Al honeycomb buffer layer \cite{mirrors}. The total photon collection area on-axis, after accounting for shadowing and other optical losses, is $50\UU{m}{2}$. The distance between the primary and the de-magnifying secondary is 8.4\U{m}, and the effective focal length is 5.6\U{m}. 

Alignment of such a complex segmented optical system is a significant challenge, and the alignment tolerances are much stricter than for simpler, single-reflector designs. Indeed, one of the main goals of the pSCT project is to test whether the optical system can meet the technical specifications. In order to achieve this, each mirror panel is mounted on an electronically controlled Stewart platform which can position the mirror with a stepper motor precision of 3\U{\mu m} over six degrees of freedom. The mirrors are equipped with edge-sensors, composed of a laser diode/ webcam pairing, which allows an initial panel-to-panel alignment and monitoring of the optical system alignment during operations. Optical tables located at the center of the primary and secondary reflectors house a laser-based global alignment system which aligns the complete primary, secondary and the camera focal plane (see \cite{pSCT_optics_icrc2015} for further description of alignment system design). The final stage of the alignment process uses a CCD image of a bright star at the focal plane to fine-tune the alignment of individual panels.  

\begin{figure}[h]
    \centering
    \includegraphics[width=\columnwidth]{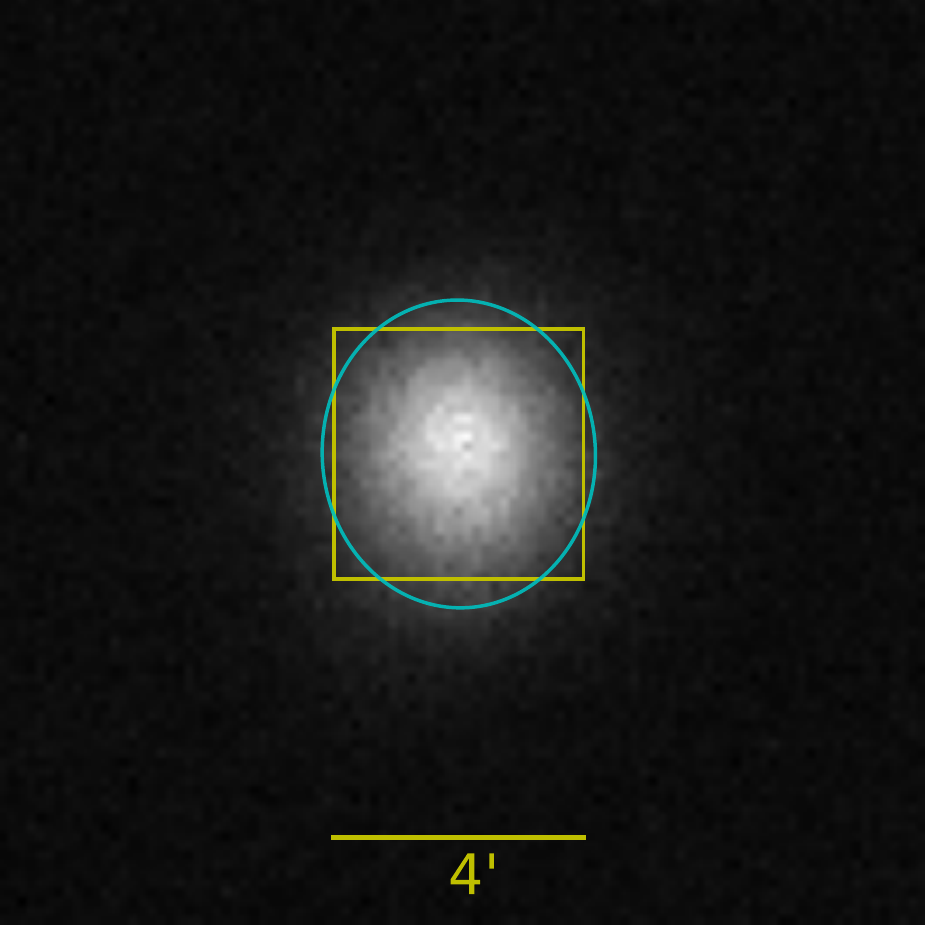}
    \captionof{figure}{An optical image of the star Capella in the focal plane of the pSCT camera recorded using a CCD camera. The cyan ellipse shows the 1.8-$\sigma$ ($\sim$80\% containment) contour from the best 2D Gaussian fit. The yellow square illustrates the size of a pSCT image pixel (an angular scale of 4 arcmin). The gray-scale indicates the light intensity, in arbitrary units.}
    \label{fig:optical_psf}
\end{figure}

A first successful alignment of the full system was completed in December 2019. Figure~\ref{fig:optical_psf} shows an image of the star Capella at the focal plane of the pSCT. The 2$\sigma$ containment radius is 2.8 arcmin, which meets the design specification of 3.6 arcmin. Almost all of the light (75\%) from a point source is contained within the 4 arcmin square Cherenkov camera image pixel. Significant improvement of the alignment is still possible with future steps, including off-axis alignment and the development of improved stability of the alignment system during telescope motion and temperature variations.

An equally important goal of the pSCT project is to demonstrate the technical feasibility of a low cost, extremely high density, high speed, modular silicon-photomultiplier camera for ground-based gamma-ray astronomy. The use of solid-state photo-detector technology for atmospheric Cherenkov telescopes was first accomplished by the FACT project \cite{FACT}, and provides significant advantages in photon detection efficiency, per-channel cost and photo-sensor density over traditional photomultiplier-tube approaches.

\begin{figure}[h]
	\centering
	    \includegraphics[width=\columnwidth]{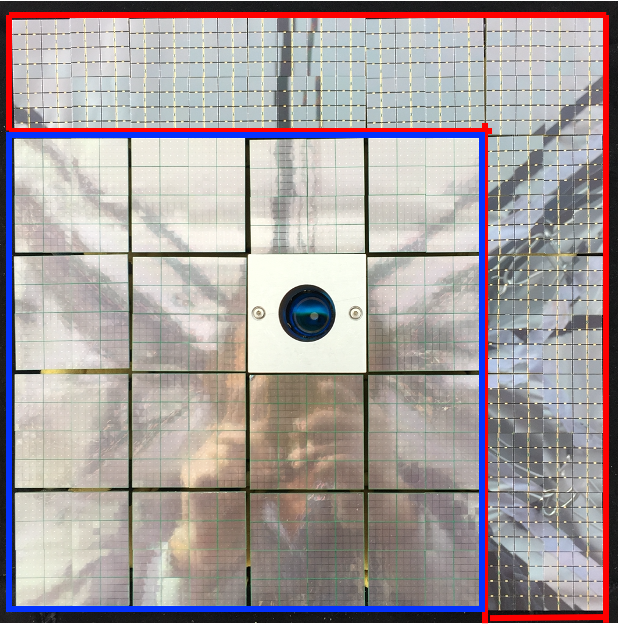}
	\captionof{figure}{The focal plane photosensor array of the current pSCT camera. The nine FBK SiPM modules are outlined in red, the remainder are Hamamatsu, outlined in blue. The central location currently houses an optical alignment module. The module pitch spacing is $54\U{mm}$, and the camera field of view subtends $2.7^{\circ}\times2.7^{\circ}$.}
	\label{fig:CurrentCameraMapping}
\end{figure}

The current pSCT camera \cite{camera_icrc2019}, installed in 2018,  represents the first stage of the project and is equipped with 24 modules in a square grid configuration. The central position contains a temporary optical alignment module, at present, as illustrated in Figure~\ref{fig:CurrentCameraMapping}. While this will unavoidably cause some truncation of gamma-ray event images from a source located at the center of the field of view, the effect should be relatively minor, since the Cherenkov emission peaks at an angular distance of approximately one degree from the source position.  

Each camera module consists of a focal plane unit housing 64 SiPM image pixels, each with a $6\U{mm}\times6\U{mm}$ photosensitive area, and a front-end electronics unit containing the associated pre-amplifier, digitization and low-level trigger electronics. Fifteen of the modules are equipped with SiPMs commercially produced by Hamamatsu (model S12642-0404PA-50(X)), and the remaining nine modules are equipped with SiPMs developed by Fondazione Bruno Kessler (FBK) in collaboration with the Istituto Nazionale di Fisica Nucleare (INFN) \cite{INFN_SiPMs1, INFN_SiPMs2, INFN_SiPMs3}. A trigger pixel, which is formed from the analog sum of the signals from four adjacent image pixels, generates a trigger when the trigger pixel signal crosses a discriminator threshold. The individual modules are connected to a custom backplane which performs camera-level trigger decisions, housekeeping and power-supply management. A camera trigger is generated by a coincidence of three adjacent trigger pixels and initiates read-out of the TARGET 7 (7th generation of the ``TeV Array Readout with GSa/s sampling and Event Trigger") digitizing application-specific integrated circuits (ASICs). More information on the design and calibration of the various generations of TARGET ASICs can be found in \cite{TARGET1, TARGET5, TARGET7}. The length of the signal waveform readout window is adjustable, and is currently 128\U{ns}, recorded in $1\U{ns}$ samples. The waveforms for all camera pixels are recorded to disk for each triggered event, and stored for offline analysis.  

Each SiPM camera image pixel views a $0.067^{\circ}\times0.067^{\circ}$ region of the sky, and the full field of view of the existing camera is $2.7^{\circ}\times2.7^{\circ}$. In the final pSCT design, the camera will house 177 modules covering an $8.0^{\circ}$ field of view with 11328 SiPM pixels. The full camera is expected to be installed and operating in 2022 \cite{camera_upgrade}.


\section{Observations and operations}

Observations of the Crab Nebula with the pSCT began on January 18, 2020, and continued until February 26, 2020. Data were collected primarily in ON/OFF mode, in which 28-minute observations of the source (ON) are preceded or followed by 28-minute observations of a blank field (OFF). The OFF-source observations are offset from the target in right ascension by an amount which ensures that they cover the same elevation angle range as the ON-source exposure. The order of each pair (ON/OFF or OFF/ON) was chosen so as to maximize the elevation (and minimize the airmass) at which the data were taken. The typical hardware trigger rate was approximately $100\U{Hz}$. In this prototype system, the majority of these triggers are due to electronic noise, with just a few Hz of actual cosmic ray events. The trigger rate also includes $10\U{Hz}$ artificially injected by uniform illumination of the camera with light pulses from an LED calibration system. Housekeeping measurements (module temperature and SiPM current readings) were taken every few minutes, briefly interrupting the data acquisition. 


After removing data with major hardware problems or poor weather (as determined by the observers on site), the total exposure, without correction for acquisition deadtime, is 21.6\U{hours} ON and 17.6\U{hours} OFF. Four hours of the ON source data do not have matching OFF source observations. We note that operating parameters were varying during these commissioning observations, during which the telescope performance characteristics were still being investigated. In particular, significant changes to the hardware trigger were implemented, including the masking of particularly noisy regions of the camera. A temperature stabilization period of $\sim5\U{minutes}$ prior to observations was also introduced, which improved system stability and noise performance. Additionally, a changing subset of typically 3 of the 64-pixel camera modules were not operating for all of the observations, due to hardware and communications problems. However, all operating parameters were held constant for the full duration of each ON/OFF pair, to reduce any chance of the changing conditions introducing a systematic bias which could mimic a signal.

\section{Analysis with the pSCT and VERITAS}
The standard approach to the analysis of imaging atmospheric Cherenkov telescope data requires the development of extensive Monte Carlo simulations of air shower development and a detailed model of the telescope optical and electronic response. These simulations are used to establish the gamma-ray selection criteria, and to calibrate algorithms used to estimate the gamma-ray primary energy \cite{iact_analysis}.  
However, the pSCT camera is an experimental prototype, with known limitations. The camera will soon be significantly revised, and substantially replaced, during the forthcoming camera upgrade. A major investment in accurate simulation, including a full description of the complex electronic noise performance, is therefore not justified. Fortunately, the fact that the pSCT is co-located with the VERITAS array means that simulations are not required, if the goal is simply to establish proof of concept by detecting an astrophysical gamma-ray source. VERITAS, co-located with the pSCT, is a well-calibrated, mature facility, which can provide independent information about the nature and properties of air showers observed simultaneously by the two instruments. In particular, it can clearly identify true air shower events (as opposed to electronic or night-sky-background noise-triggered events) and determine, with high confidence, which of these were initiated by gamma-ray primaries. 

VERITAS has been in full scientific operation at the Whipple Observatory since 2007, and its current sensitivity and performance are summarized in \cite{VERITAS_Performance}. It consists of four, $12\U{m}$ aperture Davies-Cotton telescopes separated by approximately $100\U{m}$. Each telescope is equipped with a 499-photomultiplier-tube camera, covering a field of view with a diameter of $3.5^{\circ}$. Cosmic ray air showers are recorded at a rate of $\sim400\U{Hz}$ at elevation angles above $60^{\circ}$. The data are analysed using a standard analysis chain \cite{VERITAS_analysis} which identifies gamma-rays in the $85\U{GeV}$ -- $30\U{TeV}$ range. An astrophysical source with a gamma-ray flux and spectrum similar to that of the Crab Nebula can be detected at the $5\sigma$ level in $1\U{minute}$.

\begin{figure}[h]
    \centering
    \includegraphics[width=\columnwidth]{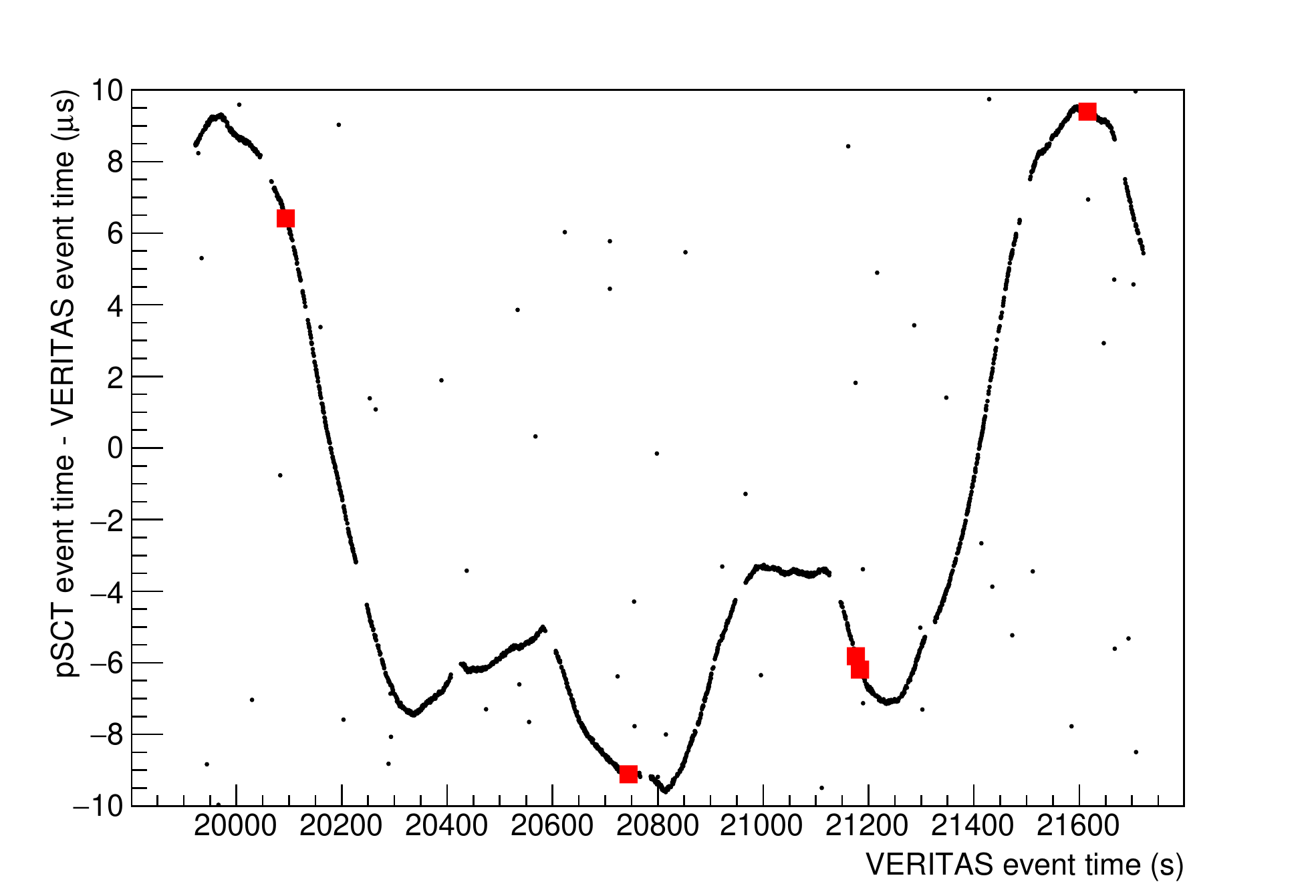}
    \captionof{figure}{The arrival time difference between events recorded by the pSCT and by VERITAS for a single Crab Nebula observation run (black points). The pSCT times have been corrected for a linear drift and a fixed offset (the drift is $0.385\U{\mu s}\UU{s}{-1}$ for this run, empirically determined using the times of well-matched cosmic ray events themselves). The width of the thick black curve in the y-direction illustrates the $\pm50\U{ns}$ coincidence. Red squares indicate gamma-ray candidate events, identified by the VERITAS analysis.}
    \label{fig:time_matching}
\end{figure}

Analysis of the pSCT Crab Nebula observations begins by establishing a dataset of clearly identified air shower events. Both VERITAS and the pSCT are equipped with high frequency oscillators ($125\U{MHz}$ for the pSCT, and $10\U{MHz}$ for VERITAS), which provide precise event timestamps. After accounting for systematic timing offsets and drifts (primarily a run-dependent linear drift term on the order of $1\U{\mu s}\UU{s}{-1}$), these timestamps can identify air showers observed simultaneously by both instruments, within a coincidence window of $\pm50\U{ns}$ (Figure~\ref{fig:time_matching}). The data sample used for this step consists of $2.2\U{hours}$ of coincident observations from January 28, 2020, recorded with a source elevation above $60^{\circ}$. A total of 11615 coincident air shower events were identified, with a typical coincident event rate of $1.5\U{Hz}$. This low rate (and correspondingly high energy threshold) is primarily a consequence of the high trigger thresholds which are required for operation of the electronic noise dominated prototype camera.

\begin{figure}[h]
    \centering
    \includegraphics[width=\columnwidth]{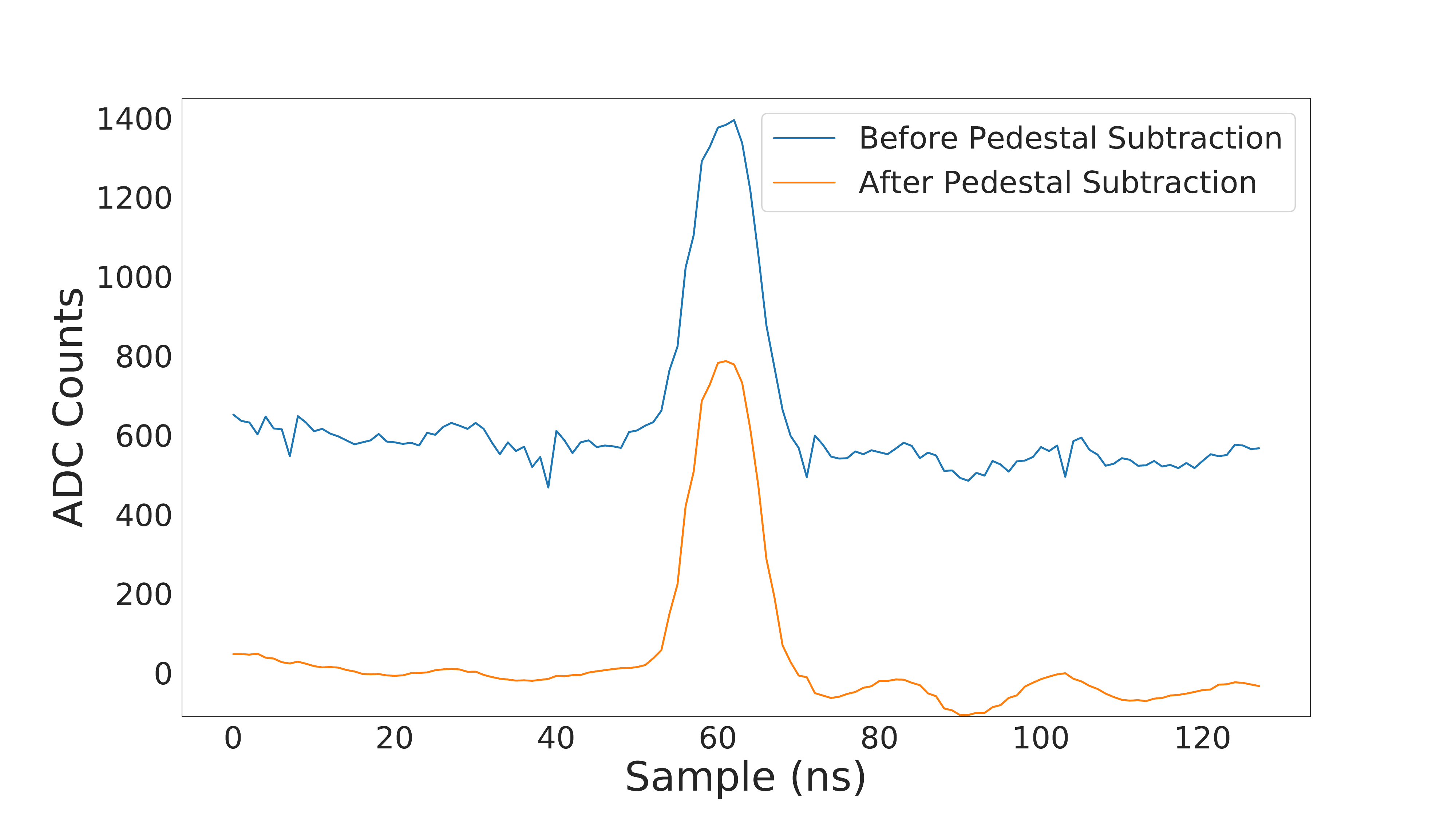}
    \captionof{figure}{We demonstrate the effect of the sample-dependent pedestal subtraction on an example signal waveform. The upper (blue) trace shows the data recorded by the camera in analog to digital converter (ADC) counts. The result of the subtraction is shown in the lower (orange) trace.}
    \label{fig:ped_sub}
\end{figure}

Working with this cosmic ray event sample, we next establish analysis procedures to process and parameterize the raw pSCT images. The pSCT data consist of a list of camera events which contain the 128-sample digitized waveforms for each of the 1536 SiPM pixels. These waveforms are calibrated by subtracting the pedestal (the response in the absence of any signal), which is tabulated as a lookup table accounting for the position of each storage cell in the switched capacitor array, as illustrated in Figure~\ref{fig:ped_sub}.

The signal for each channel is simply determined by searching the waveform to find the sample with the largest number of 
analog-to-digital converter (ADC) counts, and then integrating the pedestal-subtracted counts over a window of $\pm8\U{ns}$ centered on this peak value. An approximate conversion of the integrated ADC signal to an estimate of the number of photo-electrons generated at the SiPM photocathode is accomplished by studying the calibration LED events using the ``photostatistics" method described by Hanna et al. \cite{photostats}. We note that this is not strictly accurate, in the case of silicon photomultipliers, and future work will apply more appropriate analyses which account for excess noise due to the correlated processes of crosstalk and afterpulsing \cite{Vinogradov}.

Camera events that do not contain a clear air shower image are next identified and removed by requiring images to have at least four adjacent pixels with signals greater than 2 photo-electrons. Events triggered by the calibration LED flasher are also removed, based on their arrival times. The images for events which pass these criteria are then cleaned to remove image pixels which do not contain a clear Cherenkov signal. For this step, we use a modified version of the aperture image cleaning procedure of Wood et al. \cite{Wood_2016}, with an aperture cleaning radius equal to twice the image pixel size ($0.134^{\circ}$). In this procedure, at each pixel location, the signals from all other pixels within the aperture cleaning radius are summed together, with signals from pixels partially contained by the aperture weighted appropriately. If this summed signal crosses a threshold, the pixel on which the cleaning aperture is centered is kept for future analysis. For this work, reasonable thresholds were determined empirically for each cleaning aperture, by examining the effect of the cleaning on a sample of randomly triggered noise events which do not contain Cherenkov light. We stress that, in contrast to most operating Cherenkov telescopes, the dominant noise component in the signal traces for this prototype instrument is electronic and readout noise, as opposed to night sky background photons. 

The remaining cleaned images are then parameterized using a simple geometrical moment analysis \cite{Hillas_85}, and the resulting Hillas image parameters (\textit{size}, \textit{length}, \textit{width}, \textit{distance}, $\alpha$, etc.) are determined.

\begin{figure}[h]
    \centering
    \includegraphics[width=\columnwidth]{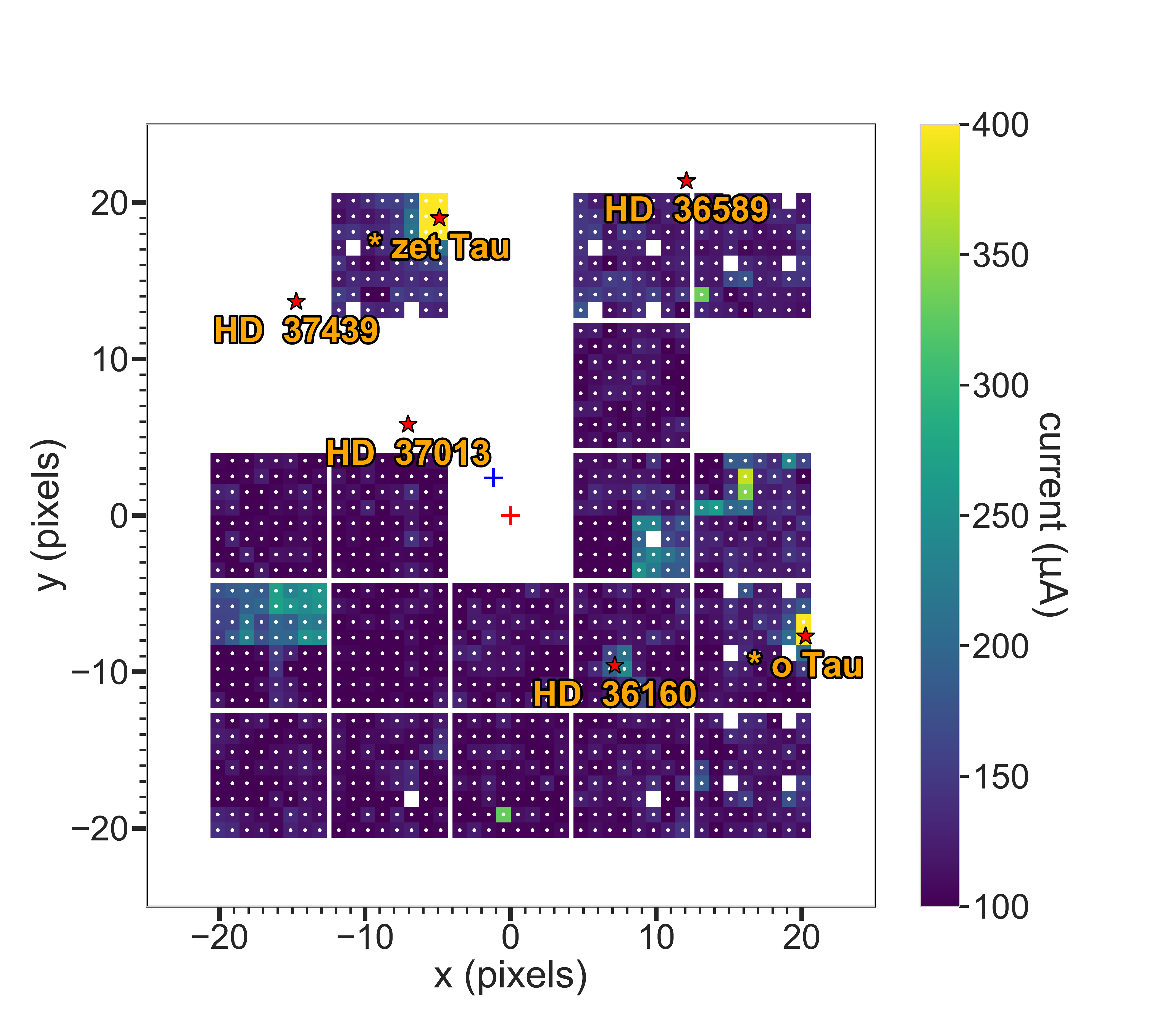}
    \captionof{figure}{Example of the pointing correction. The red cross indicates the center of the field of view, while the blue cross shows the actual position of the Crab Nebula, as derived by aligning the current image with the star field at that time. Modules and pixels with current-readout hardware problems are represented in white. 
    }
    \label{fig:pointing}
\end{figure}

\begin{figure}[h]
    \centering
    \includegraphics[width=\columnwidth]{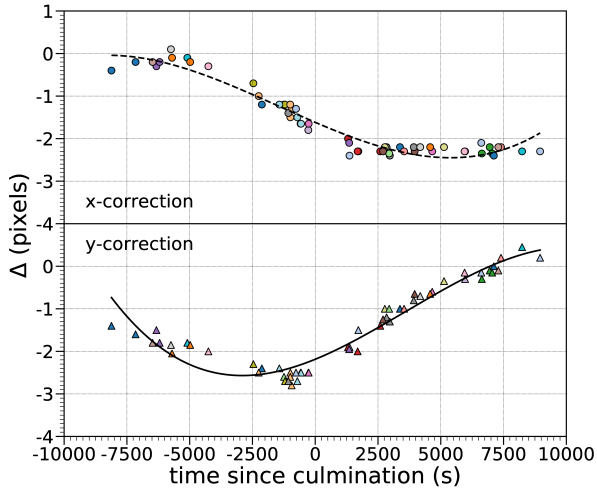}
    \captionof{figure}{Pointing correction along $x$ (top) and $y$ (bottom), in camera coordinates, as a function of time since culmination of the Crab Nebula (or its OFF source counterpart). Multiple runs from different nights are shown; points of the same colour are from the same run.The black curves are polynomial fits.}
    \label{fig:pointing_correction}
\end{figure}

Before the complete set of Hillas parameters are calculated, however, we must first correct for any errors in the telescope pointing. While the Crab Nebula (or its OFF source counterpart) was nominally at the center of the camera during these observations, the telescope bending model was not yet determined, and pointing corrections were not applied during tracking, resulting in a residual pointing offset. Offline pointing corrections for Cherenkov telescopes are typically calculated using independent CCD cameras that view the sky and the Cherenkov camera at the same time. For finely-pixellated cameras such as the pSCT, the photosensors of the Cherenkov camera itself can be used to derive pointing corrections with a similar level of accuracy, using the locations of stars within the field of view \cite{CANGAROO}. In this analysis, the star locations are measured using the SiPM anode currents, as illustrated in Figure~\ref{fig:pointing}. During each run, the SiPM currents are read out every 3 minutes. For this analysis, we first measured the pointing offset using a few current maps taken from each run, and then calculated the offset at a given time using a linear fit to these offsets as a function of telescope elevation. The maximum calculated offsets are typically on the order of a few SiPM pixels ($\sim0.2^{\circ}$), as illustrated in Figure~\ref{fig:pointing_correction}. The true position of the Crab Nebula in the field of view is then used to derive the Hillas parameters for each image. This includes $\alpha$, defined as the angle between the major axis of the image and a line joining the image centroid to the putative source position.


\begin{figure*}[h]
    \centering
    \includegraphics[width=1.0\textwidth]{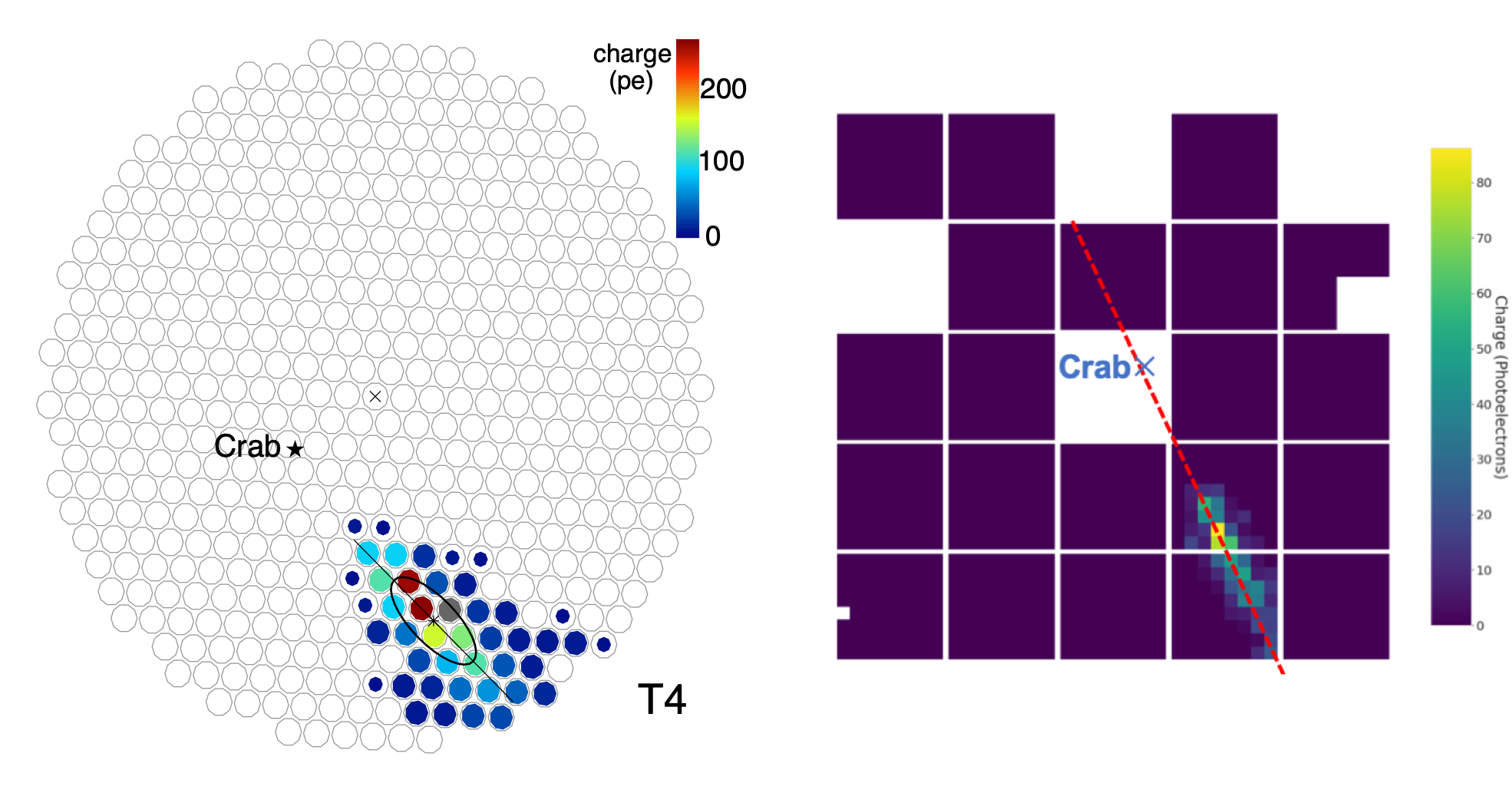}
    \captionof{figure}{The same air shower event, observed by VERITAS Telescope 4 (left) and the pSCT (right). VERITAS array reconstruction identifies this as a $3.5\U{TeV}$ gamma-ray candidate. The angular scale of the two images is the same. The image on the right shows the cleaned pSCT image, together with missing or non-functioning modules (white space). The true position of the Crab Nebula in the field of view, after pointing corrections, is also indicated.  } 
    \label{fig:camera_image}
\end{figure*}

The final stage of the analysis concerns the separation of gamma-ray initiated events from the cosmic ray background. In addition to identifying cosmic ray air shower events, the standard VERITAS reconstruction tools can be used to establish, with high confidence, which events are likely to be due to incident gamma-rays. For the $2.2\U{hour}$ January $28^{\mathrm{th}}$ data sample, 18 coincident events passed the VERITAS gamma-ray selection cuts. Figure~\ref{fig:time_matching} illustrates the arrival times of a subset of these gamma-ray candidates, while Figure~\ref{fig:camera_image} shows one of these events as seen by both the pSCT, and by VERITAS Telescope 4. Telescope 4 is closest to the pSCT, at a distance of only $35\U{m}$, and therefore has a very similar view of the Cherenkov emission from most air showers. 

\begin{figure}[h]
    \centering
    \includegraphics[width=\columnwidth]{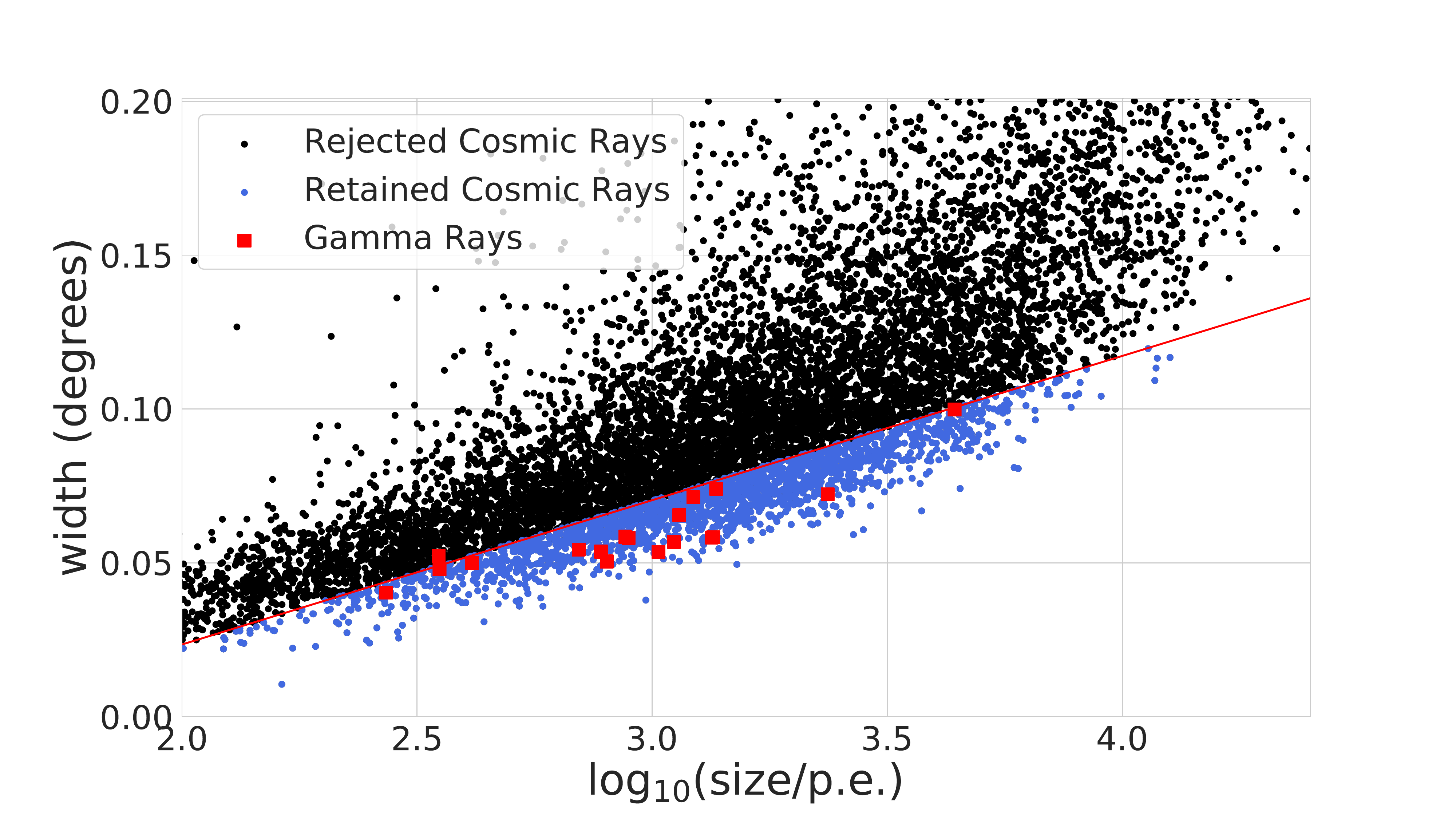}
    \includegraphics[width=\columnwidth]{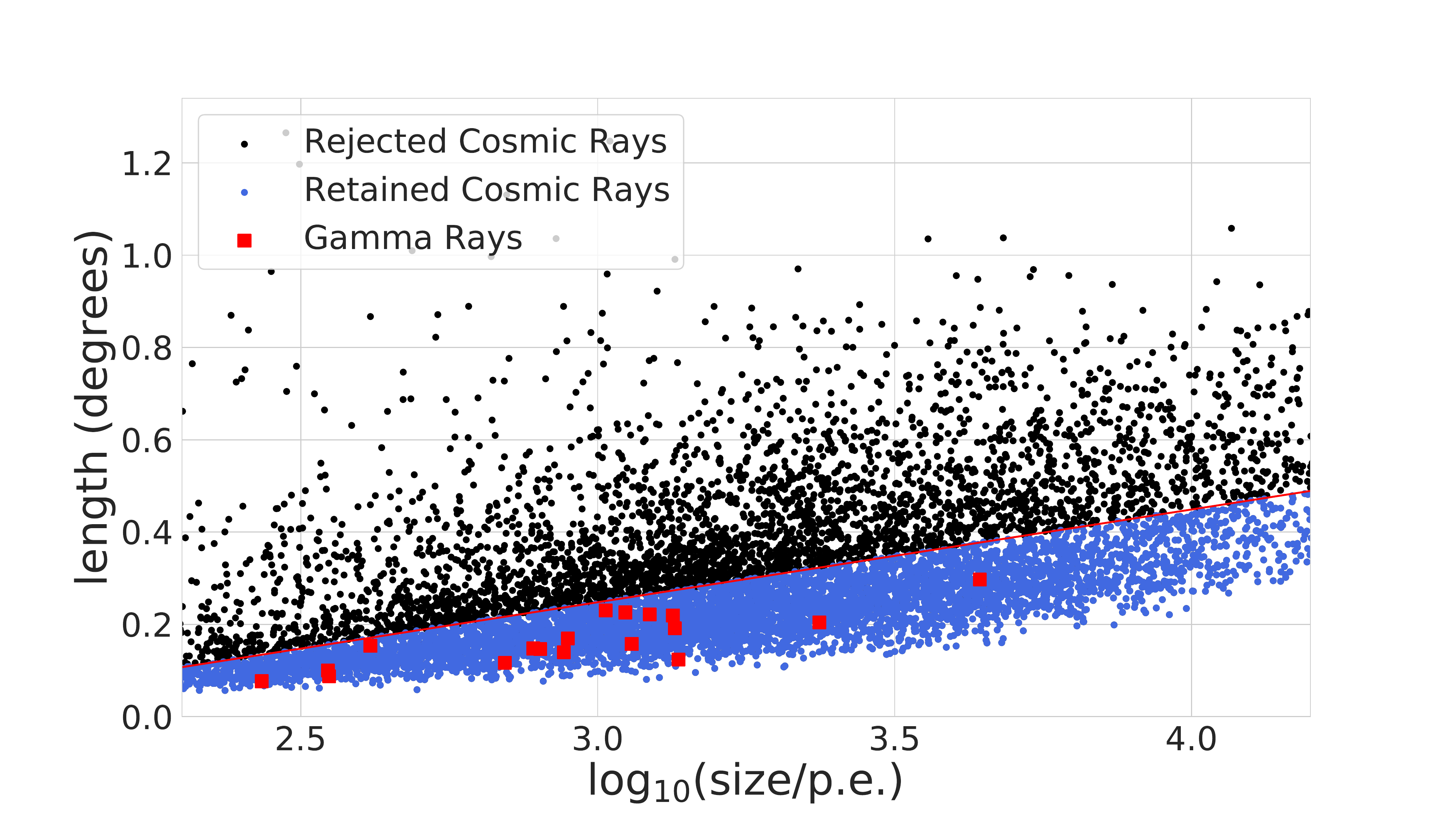}
    \captionof{figure}{The Hillas \textit{width} (top) and \textit{length} (bottom) parameters, as a function of image \textit{size}, for the $2.2\U{hour}$ sample of overlapping VERITAS and pSCT data. The gamma-ray selection cuts in Table~\ref{tab:first_cuts} retain events below the red lines. Retained events are shown in blue, while rejected events are in black. Gamma-ray candidates identified by VERITAS are indicated by red squares. 
    }
    \label{fig:crab_width_v_size}
\end{figure}

This sample of 18 gamma-ray candidate events, and 11597 cosmic ray events, can then be used to optimize the pSCT image parameter selection cuts which provide discrimination between the two populations. The parameter selection cuts we use are based on those originally developed for the Whipple 10\U{m} telescope \cite{Supercuts, Ext_Supercuts}. Figure~\ref{fig:crab_width_v_size} illustrates the \textit{size}-dependent cut boundaries for the image \textit{width} and \textit{length} parameters. For this first analysis all cuts were optimized manually, and constrained to retain $\sim{95\%}$ of the very limited gamma-ray sample after all cuts were applied. The full set of optimized cuts is detailed in Table ~\ref{tab:first_cuts}. These cuts are then applied to the full remaining sample ($17.6\U{hours}$) of ON/OFF observations of the Crab Nebula with the pSCT. We stress that the data used to derive the cuts are excluded from this final stage, as are any ON source observations without matching OFF source data.

\begin{table}\centering
\caption{Gamma-ray selection cuts optimized using VERITAS matched events.}
\label{tab:first_cuts}
\begin{tabular}{c} \toprule
Gamma-ray selection cuts \\  \midrule
    $0.33^{\circ} < distance < 1.14^{\circ}$\\
    $250\U{p.e.} < size < 20000\U{p.e.}$\\
    $width < -0.070^{\circ}+0.047^{\circ}\log_{10}(size/p.e.)$ \\
    $length < -0.369^{\circ}+0.201^{\circ}\log_{10}(size/p.e.)$\\
    $1.139^{\circ}-1.742^{\circ}(width/length) < distance$\\
    $distance < 1.273^{\circ}-0.737^{\circ}(width/length)$\\
        $\alpha < 6^{\circ}$\\
 \bottomrule
\end{tabular}
\end{table}

\begin{figure}[h]
    \centering
    \includegraphics[width=\columnwidth]{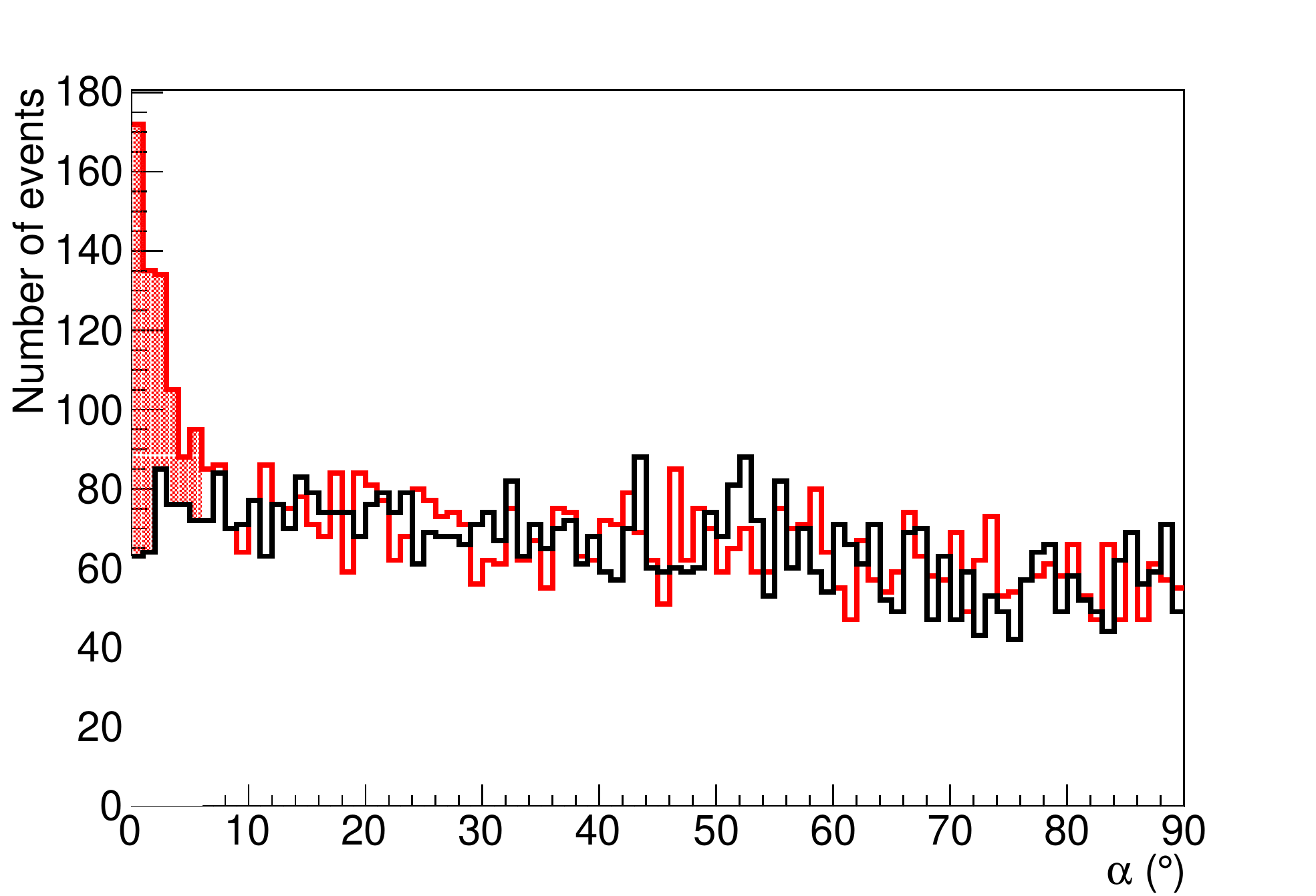}
    \captionof{figure}{$\alpha$: the angle between the major axis of the image, and a line joining the centroid of the image to the location of the Crab Nebula. The red histogram is for $17.6\U{hours}$ of ON source observations, black is for the same duration of OFF source observations, after applying the gamma-ray selection cuts in Table~\ref{tab:first_cuts}. The shaded region indicates the $<6^{\circ}$ cut on $\alpha$ itself. 
    }
    \label{fig:crab_alpha}
\end{figure}

Figure~\ref{fig:crab_alpha} shows the distribution of the $\alpha$ parameter for both ON and OFF source observations. An ON-source excess is apparent at low values of $\alpha$, corresponding to a statistical significance of over 8 standard deviations ($\sigma$, calculated using equation 17 of \cite{LiMa}), providing a clear detection of gamma-ray emission from the Crab Nebula. The average gamma-ray rate is $0.28\pm0.03\UU{min}{-1}$. As with the cosmic ray rate, this gamma-ray rate is very low, when compared with the expectation based on the Crab results from single telescopes with similar photon collection efficiencies (e.g. a few events per minute, for the Whipple 10m). This is consistent with the conclusion that the current pSCT energy threshold is much higher than the ultimate design goal, primarily as  a result of the electronic noise dominated trigger threshold. 

\begin{figure}[h]
    \centering
    \includegraphics[width=\columnwidth]{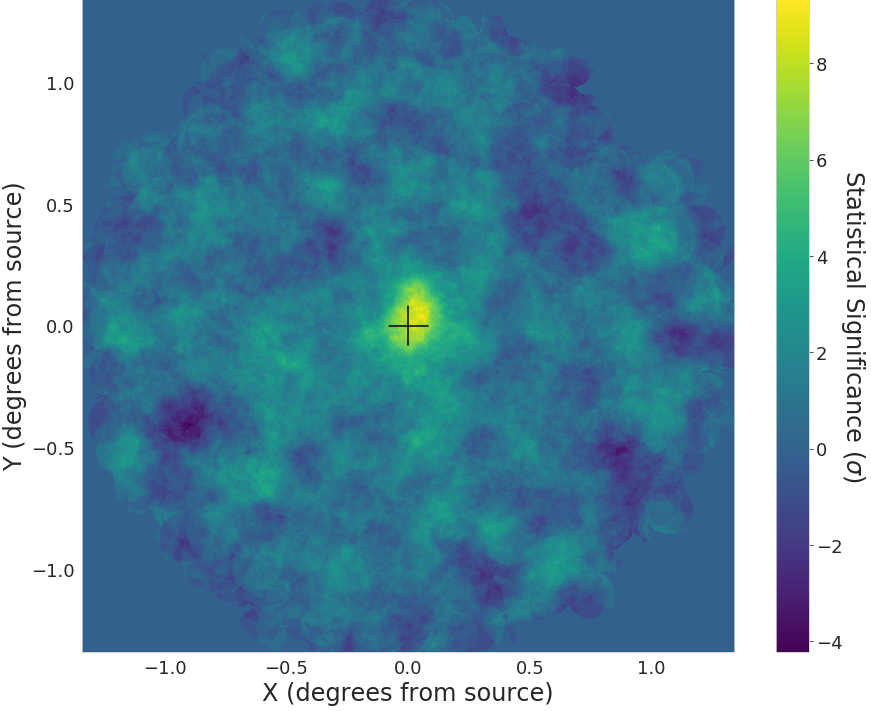}
    \captionof{figure}{A significance sky map of the Crab Nebula observations in camera coordinates. The cuts in Table~\ref{tab:first_cuts} (except for the cut on the  Hillas $\alpha$ parameter) are used to select events from the $17.6\U{hours}$ of matched ON and OFF source data. Shower arrival direction is reconstructed using the \textit{disp} method \cite{Lessard}. Top hat smoothing with a smoothing radius of $0.1^{\circ}$ is then applied.}
    \label{fig:crab_skymap}
\end{figure}

The Crab Nebula detection is further illustrated in Figure~\ref{fig:crab_skymap}, which shows a 2-D map of the gamma-ray emission generated using the method of \cite{Lessard}. The arrival direction of each air shower is derived by estimating the angular distance along the long axis of the image between the image centroid and the source position using the \textit{disp} parameter, which is defined as: 
\begin{equation}
        disp = \xi(1 - \frac{width}{length})
\label{equation:disp}
\end{equation}

The constant $\xi$ was empirically determined from the mean of the $distance$ distribution for the gamma-ray candidate events identified by VERITAS ($\xi = 1.24^{\circ}$). The skymap is then constructed by calculating the statistical significance of each point in the map, using the number of ON and OFF events which lie within $0.1^{\circ}$ of the selected map point. 

\section{Analysis with the pSCT alone}
The analysis described above was optimized using just 18 gamma-ray candidate events, and a small ($2.2\U{hour}$) sample of overlapping VERITAS and pSCT data. This allowed a clear detection of a strong gamma-ray source. However, with this reasonably strong detection in hand, it is possible that the sensitivity of the analysis can be improved by optimizing on a larger sample of pSCT data alone. 

To explore this possibility, we divide the $17.6\U{hours}$ of ON/OFF data into two subsets: a ``training" sample, consisting of $5.9\U{hours}$ of observations spread throughout the campaign, and a ``test" sample of the remaining $11.7\U{hours}$. Applying the analysis described in the previous section to the training sample results in a signal with a significance of $5.7\sigma$. Re-optimizing the analysis using only this training sample results in the gamma-ray selection cuts shown in Table~\ref{tab:second_cuts}. The re-optimization was conducted simply by sequentially changing the individual cut parameters and observing the effect on the significance and the excess rate of the training sample. The final set of cuts will therefore not be perfectly optimized: this would require a full multi-dimensional minimization. However, they represent an improvement over the original analysis, providing an \textit{a posteriori} significance of $7.2\sigma$ when applied to the sample of data on which they were trained.  Applying these cuts \textit{a priori} to the test sample improves the statistical significance for this sample from $6.7\sigma$ to $7.3\sigma$. Probably more important than this modest improvement is that this demonstrates that pSCT data alone can be used to test and further refine the analysis in future, without requiring strictly coincident VERITAS observations. A full summary of the results is given in Table~\ref{tab:results}.



\begin{table}\centering
\caption{Gamma-ray selection cuts optimized using the pSCT training sample.}
\label{tab:second_cuts}
\begin{tabular}{c} \toprule
Gamma-ray selection cuts \\  \midrule
    $0.47^{\circ} < distance < 1.07^{\circ}$\\
    $250\U{p.e.} < size < 20000\U{p.e.}$\\
    $width < -0.072^{\circ}+0.045^{\circ}\log_{10}(size/p.e.)$ \\
    $length < -0.342^{\circ}+0.201^{\circ}\log_{10}(size/p.e.)$\\
        $\alpha < 6^{\circ}$\\
 \bottomrule
\end{tabular}
\end{table}
\begin{table}\centering
\caption{A summary of the analysis results. ``$t$" is the ON-source exposure time in hours.}
\label{tab:results}
\begin{tabular}{lcccc} \toprule
\multicolumn{5}{c}{Cuts optimized using VERITAS matching} \\
\midrule
     & ON & OFF & $\sigma$ & $\sigma/\sqrt{t}$\\
    Full dataset & 729 & 436 & 8.6 & 2.05\\
    Training sample & 202 & 104 & 5.7 & 2.35\\
    Test sample & 527 & 332 & 6.7 & 1.96\\
    \midrule 
\multicolumn{5}{c}{Cuts optimized using the pSCT training sample} \\
    \midrule
    Test sample & 307 & 152 & 7.3 & 2.13\\
     \bottomrule
\end{tabular}
\end{table}
\section{Conclusions and prospects}

The results presented here are a snapshot of the current commissioning status of the pSCT. While they are encouraging, and represent an important milestone in the project development, it is clear that the telescope is currently operating far below its ultimate design sensitivity. The reasons for this were foreseen, for this experimental prototype, and are reasonably well understood. In particular,  both the online triggering behavior and the offline waveform and image analysis performance are dominated by electronic noise, as opposed to night-sky-background light. This issue will be addressed with the upgrade to the full, $8^{\circ}$ field of view camera, which is currently in production. Critically, the upgraded camera will use a revised version of the TARGET chips, in which the independent trigger and digitization functions are implemented in two separate ASICs (T5TEA and TARGET C, respectively \cite{TARGETC}), to remove electronic noise coupling between the two paths. It will also include improved pre-amplification, SiPM bias voltage control and pulse-shaping through the use of the SMART ASIC. A physical redesign of the camera electronics modules will improve reliability and shielding, and further minimize cross-talk and noise pickup on the analog lines. Laboratory tests indicate that these modifications are effective, reducing the electronic noise to the single digital count level. 

The pSCT will continue to operate while the upgraded camera is under construction. The F.L. Whipple Observatory
re-opened in October 2020, following its temporary closure in March 2020 due to the COVID-19 epidemic. An intensive program of pSCT observations and engineering tests are planned for 2020-2021. These include work on the optical alignment, to ensure stable and optimum optical performance over the full field of view and over all azimuth and elevation angles. The telescope's mechanical pointing model will also be tuned, and various improvements to the camera hardware, operational software and observing procedures will enhance reliability and stability.

Offline software development work is required to build accurate Monte Carlo simulations of the telescope response which will allow to measure gamma-ray source fluxes, energy spectra and morphology. Analysis tools which take full advantage of the pSCT's high resolution optics and camera are also under development. As experience with prior generations of Cherenkov telescopes shows, the ability to rapidly and consistently detect a strong astrophysical gamma-ray source is invaluable to such efforts. The Crab Nebula can be easily detected by the pSCT in a short exposure ($5.5\U{hours}$ of ON source observations for a $5\U{\sigma}$ detection, using the cuts in Table~\ref{tab:second_cuts}). The analysis and simulation development of the pSCT will be further boosted by continued coordinated observations with VERITAS, and the pSCT is expected to eventually operate as an additional telescope in the VERITAS array, providing a significant enhancement in sensitivity and performance. 

Ultimately, however, the pSCT's primary purpose is to serve as a pathfinder for a major contribution to CTA. The goal of the SCT team is to deliver at least 10 SCTs to the CTA observatory, which will considerably enhance the overall performance of CTA. Much of the pSCT effort over the next few years will be in support of this goal, including a focus on developing the manufacturing and systems engineering required for an SCT contribution to this major astronomical observatory. The timely addition of SCTs will help CTA to play a leading role in multimessenger astrophysics, and to achieve its wide-ranging key science goals: studying astrophysical particle acceleration, probing extreme environments in the Universe and addressing fundamental physical questions, including the nature of dark matter \cite{cta_science}.

\section{Acknowledgments}

This research is supported by grants from the U.S. National Science Foundation and the Smithsonian Institution, by the Istituto Nazionale di Fisica Nucleare (INFN) in Italy and by the Helmholtz Association in Germany. The development, construction and operation of the pSCT was supported by NSF awards 
(PHY-1229792, PHY-1229205, PHY-1229654, PHY-1913552, PHY-1807029, PHY-1510504, PHY-1707945, PHY-2013102, PHY-1707544, PHY-2011361, PHY-1707432, PHY-1806554) together with funds from
Barnard College,
California State University East Bay,
Columbia University,
Georgia Institute of Technology,
Iowa State University,
Smithsonian Institution,
Stanford University,
University of Chicago,
University of Alabama in Huntsville,
University of California,
University of Iowa,
University of Utah,
University of Wisconsin--Madison, and
Washington University in St. Louis.
ASTRI participation in this effort was supported by the Italian Ministry of University and Research (MUR) with funds specifically assigned to the Italian National Institute of Astrophysics (INAF) for the development of technologies toward the implementation of CTA. 
Support from the Japan Society for the Promotion of Science was provided by KAKENHI grant numbers JP23244051, JP25610040, JP15H02086, JP16K13801, JP17H04838 and JP18KK0384. This work was also partially supported by UNAM-PAPIIT IG101320.
We acknowledge the excellent work of the technical support staff at the Fred Lawrence Whipple Observatory and at the collaborating institutions in the construction and operation of the instrument. We also thank the VERITAS Collaboration for their cooperation in obtaining joint observations and for the use of their data. 

\end{document}